\documentclass[dvips]{aa}
\usepackage[dvips]{graphics}
\begin{document}

\title{The hard X-ray properties of the Seyfert nucleus in NGC 1365}
\titlerunning{The hard X-ray properties of NGC1365}

\author{G. Risaliti\inst{1}, R. Maiolino\inst{2},
L. Bassani\inst{3}}
\authorrunning{Risaliti et al.}

\institute{
Dipartimento di Astronomia e Scienza dello Spazio,
Universit\`a di Firenze, Largo E. Fermi 5, I--50125 Firenze, Italy
(risaliti@arcetri.astro.it)
\and
Osservatorio Astrofisico di Arcetri, Largo E. Fermi 5,
I--50125 Firenze, Italy (maiolino@arcetri.astro.it)
\and
Istituto T.E.S.R.E.,  CNR, via Gobetti 101,
I-40129 Bologna, Italy (bassani@tesre.bo.cnr.it)
}

\offprints{G. Risaliti}

\date{}

\thesaurus{03(11.09.1; NGC 1365; 11.01.2; 11.19.1; 13.25.2}

\maketitle

\begin{abstract}

We present BeppoSAX observations of the Seyfert 1.8 galaxy NGC1365 in the
0.1--100 keV range.
The source was 6 times brighter than during an ASCA observation 
3 years earlier. The 4--10 keV flux is highly variable during
the BeppoSAX observation, while the soft (0.1-4 keV) emission is
constant within the errors.
Both a cold and a warm reflector and a cold absorber are required to explain
the observed spectrum. The comparison between ASCA and BeppoSAX spectra
strongly suggests that the circumnuclear material has a more complex
structure than a simple homogeneous torus, with quite different absorbing
gas columns along different lines of sight.
A broad iron K$_\alpha$ line is also present in the spectrum, with the peak
energy significantly redshifted. This can be explained by means of a
relativistic disk line model. Alternatively, a warm
absorption Fe line system with N$_H \simeq 10^{23}$ cm$^{-2}$ could account
for the observed line profile.
\end{abstract}

\section{Introduction.}
NGC 1365 is a barred spiral galaxy (Hubble type SB0) in the Fornax
cluster that hosts an active nucleus whose optical spectrum shows weak
broad Balmer lines (Seyfert 1.8, Alloin et al. 1981)

In this paper we present the analysis of the spectrum of NGC 1365 in the
0.1-100 keV spectral range obtained with 
the BeppoSAX satellite (Boella et al. 1997).

During the past ten years NGC 1365 has been
observed several times in the X-rays 
by ASCA (Iyomoto et al. 1997, hereafter I97)
ROSAT (Komossa \& Schulz 1998) and Ginga (Awaki 1991).
The 1-10 keV continuum spectrum observed by ASCA in August 1994 and
January 1995
(I97) is well reproduced by a flat powerlaw (photon index $\Gamma$=0.8)
and a thermal soft component. A strong emission feature is present
at E$\sim$6.4--7 keV, which can be fitted by a single
broad emission line with E=6.58 keV and equivalent width 
EW=2.1 keV or, alternatively, by two narrow lines with
E=6.4 keV (neutral iron, EW=0.9 keV) and E=6.7 keV (highly ionized iron,
EW=0.9 keV).
Both these spectral features and the lack of (short term)
variability suggested that the ASCA spectrum is dominated by a cold
reflection component which is usually observed in most of the heavily
absorbed, Compton thick sources (Maiolino et al. 1998, hereafter M98)
and generally ascribed to the reflection from the molecular torus expected
by the unified model of AGNs (Antonucci 1993).

The ASCA--SIS and ROSAT--HRI data, obtained in 1994 and 1995, reveal also
the presence of a strong off-nuclear X-ray source characterized by a
steep powerlaw spectrum (photon index $\Gamma$=1.7 in the 1-10
keV band) and by a strong variability on time-scales of months; during the ASCA observation in
1995 this source was as bright as the Seyfert 2 nucleus with a flux of 0.9$\times 10^{-12}$ erg
cm $^{-2}$ s$^{-1}$.
The spatial resolution of the BeppoSAX instruments does not allow to
separate the contribution of this source from that of the nucleus. 
We will 
 discuss the possible contamination from this off-nuclear source further in
Sect. 2.

In the next section we present the results of the spectral and temporal
analysis of our data. In Sect. 3 we discuss 
the BeppoSAX data and their differences with respect to the previous
X-ray observations.
We assume a distance of 18.4 Mpc for NGC 1365, as
estimated by Fabbiano et al. (1992),
and in agreement with more recent Cepheid
measurement (Madore et al. 1998).
\section{Data analysis}
NGC 1365 was observed by SAX in August 1997. The effective
on--source integration
time was 8900 seconds for the LECS instrument (0.1-10 keV), 30000
seconds for the MECS (1.65-10.5 keV) and 14000 seconds for the PDS
(15-200 keV).
The spectrum and the light curve of the LECS and MECS
were obtained from
the ``event files'' provided by the BeppoSAX SDC, using the standard software
for X-ray analysis FTOOLS 4.0. The PDS spectrum was obtained by the FOT
files of the SAX observation, using the XAS code, a software developed
specifically for the reduction and analysis of the SAX data.

We adopted the standard data reduction for the BeppoSAX spectra
as described, for instance, in M98. 
The final spectrum was rebinned to contain at least 20 counts/bin, so that
a gaussian statistics can be used to fit
the models to the data.\\

\subsection{Spectral analysis}

The beam-size of the PDS
($\sim 1.3^{\circ}$ FWHM) includes also the
Seyfert 2 galaxy NGC1386, also observed by BeppoSAX (M98),
that should contribute significantly to the 20--100 keV flux measured for
NGC1365 (probably up to 50\%). This problem, along with other effects observed
in the light curve (Sect.2.2), prevent us from using the
PDS data to constrain the spectral properties of the source.

The best fit to the LECS and MECS data is obtained by means of
a multi-component model typical of Compton-thin sources (see M98 for
details). The continuum emission is well reproduced by 
a powerlaw of photon index $\Gamma$=1.93 (which is typical for
Seyfert 1 spectra), a photoelectric cut-off, corresponding to a column
density of cold absorbing material N$_H\sim4\times 10^{23}$ cm$^{-2}$, and a
second powerlaw that fits the soft excess which may be due to
extended components (starburst or hot gas in the Narrow Line Region)
 or to the X-ray source resolved by ASCA and ROSAT. 
The whole spectrum is also absorbed by a Galactic column density of
$1.4\times 10^{20}$cm$^{-2}$.
If the extended contribution is dominant,
we would expect that a Raymond--Smith
model also fits well the soft data.
Unfortunately the statistics of our data in the soft band is not high enough to 
discriminate
 between a powerlaw and a thermal spectrum: a Raymond-Smith model
with kT$=2^{+0.6}_{-0.4}$ gives
 a slightly worse fit ($\Delta \chi^2 =2$) than
the powerlaw, but still in agreement at a 90\% confidence level.

In addition to these continuum components, a narrow emission line with E=6.257
keV\footnote{E=6.29 keV rest frame.} 
is strongly requested by the fit ($\Delta \chi^2$=18). Note that the
line width parameter was not frozen to zero, therefore the narrowness of
the line is a result of the fit. The line
equivalent width is EW=330$^{+70}_{-130}$ eV with respect to the observed
continuum (EW=190$^{+45}_{-75}$ eV with respect to the unabsorbed
powerlaw component).
Finally, with a second line at E=6.95 keV\footnote{E=7.0 rest frame.}
(corresponding to H-like iron) the fit is better at a level of
confidence higher than 90\% ($\Delta \chi^2$ =2.9).
We note that the energy of the cold line is significantly lower than
the value of the neutral iron K$_\alpha$ line, which is
E=6.365 keV, when corrected for the redshift (the best fit with the line
energy frozen at E=6.365 is worse by $\Delta \chi^2$ =3.5). This issue
will be discussed further in Sect. 3.

The results of our fit are summarized in Table 1 and shown in Fig. 1.\\
The fit of the low--state spectrum is not statistically good ($\chi^2=47$
for 38 degrees of freedom), but this is due to the lower
signal--to--noise of the data. As shown in Fig. 1, there
are no significant continuum features that are not well fitted, while
the high
$\chi^2 / d.o.f.$ is due to the large scatter of the points in the
2-4 keV and 8-10 keV bands.

\begin{figure}
\centerline{
\resizebox{\hsize}{!}{\includegraphics{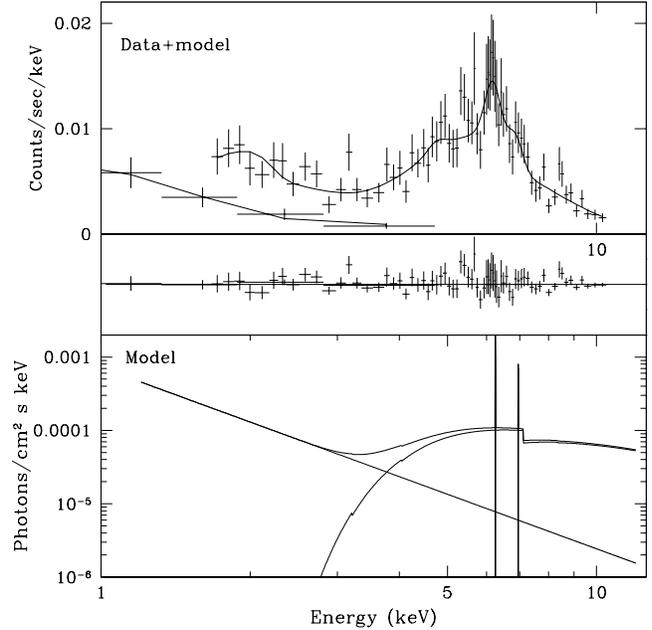}}}
\caption{{Data + model for NGC 1365 (upper panel), residuals (central
panel) and model for the best fit of our data}}
\end{figure}
\begin{table}
\centerline{\begin{tabular}{cccc}
\hline
Parameter&Best-fit value\\
\hline
&\\
\multicolumn{2}{c}{\bf Total Spectrum}\\
Powerlaw 1 $\Gamma$ & 1.93$^{+0.25}_{-0.15}$\\
Powerlaw 1 norm. & 7.4$^{+1.2}_{-0.65} 10^{-3}$ Ph. cm$^{-2}$ s$^{-1}$
keV$^{-1}$\\
N$_H$ & 4.0$^{+0.4}_{-0.5} \times 10^{23}$ cm$^{-2}$\\
Powerlaw 2 $\Gamma$ & 2.46$^{+0.40}_{-0.30}$ \\
Powerlaw 2 norm. 2 &6.8$^{+2.6}_{-2.1} 10^{-4}$ Ph. cm$^{-2}$
s$^{-1}$ keV$^{-1}$\\
Gaussian 1 Energy & 6.257$^{+0.09}_{-0.09}$ keV\\
Gaussian 1 norm. & 3.4$^{+1.3}_{-1.4} 10^{-5}$ Ph. cm$^{-2}$
s$^{-1}$
keV$^{-1}$ \\
Gaussian 1 EW & 330$^{+70}_{-130}$ eV \\
Gaussian 2 Energy & 6.95 keV (fixed)\\
Gaussian 2 norm. & 1.4$^{+1.1}_{-1.3} 10^{-5}$ Ph. cm$^{-2}$
s$^{-1}$
keV$^{-1}$ \\
Gaussian 2 EW & 120$^{+100}_{-100}$ eV \\
Flux 2-10 keV & 6.6$\times 10^{-12}$ keV cm$^{-2}$ s$^{-1}$\\
$\chi^2$/d.o.f. & 67/67 \\
\hline\\
\multicolumn{2}{c}{\bf Low state}\\
Normalization 1 & 5.8$^{+0.4}_{-0.5} 10^{-3}$ Ph. cm$^{-2}$ s$^{-1}$\\
Flux 2-10 keV & 5.0$\times 10^{-12}$ keV cm$^{-2}$ s$^{-1}$\\
$\chi^2$/d.o.f. & 47/38\\ 
\hline\\
\multicolumn{2}{c}{\bf High state}\\
Normalization 1 & 1.0$^{+0.1}_{-0.05} 10^{-2}$ Ph. cm$^{-2}$ s$^{-1}$\\
Flux 2-10 keV & 8.2$\times 10^{-12}$ keV cm$^{-2}$ s$^{-1}$\\
$\chi^2$/d.o.f. & 51.7/55 \\
\hline
\end{tabular}}
\caption{\footnotesize{Best fit model for NGC 1365. Errors are quoted
at 90\% level of confidence, using a $\chi^2$ statistics.
The model also
includes a Galactic absorbing column density of $1.4\times 10^{20}$ cm$^{-2}$.
The normalization of the powerlaws is at 1 keV. }}
\end{table}
\begin{figure}
\centerline{
\resizebox{\hsize}{!}{\includegraphics{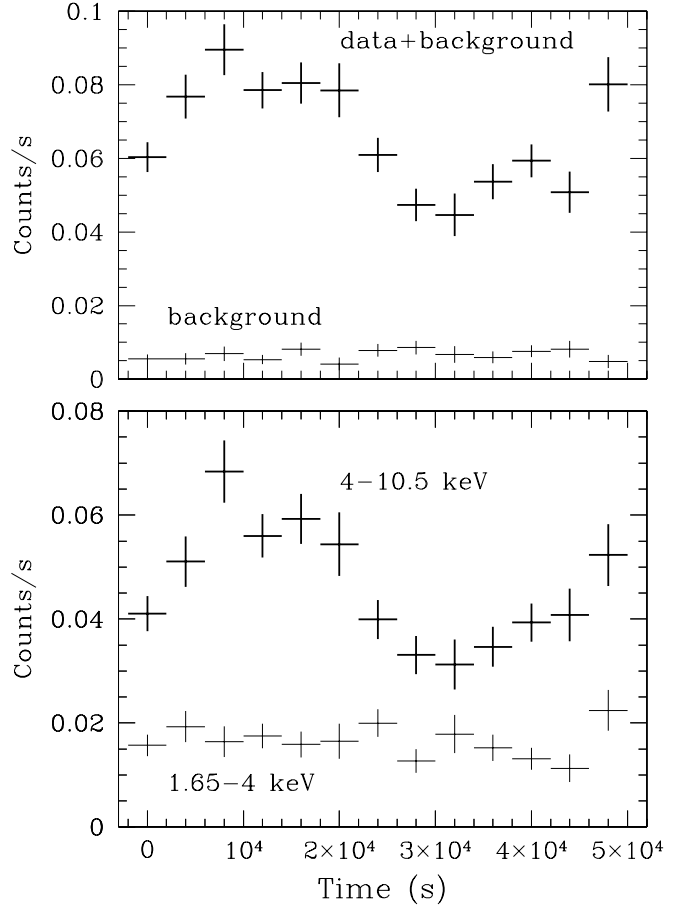}}}
\caption{{Upper panel:
light curve of the MECS (1.65-10.5 keV) observation of NGC
1365. Lower panel: light curves in the hard (4-10.5 keV) and soft (1.65-4 keV)
bands.}}
\end{figure}~

\subsection{Timing analysis}

The flux measured by BeppoSAX in the 2--10 keV band
is 6.6$\times 10^{-12}$ erg cm$^{-2}$ s$^{-1}$, about 6
times higher than the flux measured by ASCA in 1994--1995, but
similar to the flux of 4.8$\times 10^{-11}$ erg cm$^{-2}$ s$^{-1}$
measured by Ginga during manouvering operations prior to 1990 (Awaki
1991).

The light curve of NGC 1365, obtained from the MECS data (1.65-10.5 keV),
is plotted in Fig. 2.
The count rate varies by a factor of $\sim$ 2 during our observation.
There is an indication of periodicity with period
T$\simeq$ 45000 s, but longer observations are required to test
this hypothesis.

In Fig. 2 we also plot the low and high energy part of the light curve
separately. These two curves clearly show that the observed variability 
is mostly due to the high energy part of the spectrum,
while the emission in the soft part of the spectrum is
roughly constant\footnote{ This is also supported by the timing
analysis of the LECS data: in this case
the statistics is lower than in the MECS data,
however the light curve is constant
within the errors, likewise to the low--energy light curve of the MECS.}.
Comparing these results with the spectral model in Table 1, we
can conclude that the variability is due to the direct emission (above the
photoelectric cutoff) from the
central source, while the reflected or diffuse component
 (or the off-nuclear component) does not appear to vary.

We extracted two spectra from our MECS and PDS data, by selecting
the time intervals in which the
count rate in the 2-10 keV band
is respectively higher and lower than the average.
The measured fluxes in the 2-10 keV band are 8.2$(\pm 0.3)\times 10^{-12}$
erg cm$^{-2}$ s$^{-1}$ in the high state and 5.0 $(\pm 0.2)\times 10^{-12}$
erg
cm$^{-2}$ s$^{-1}$ in the low state.
The LECS and MECS spectra ($\sim$ 1-10 keV) of the
high- and low-state spectra can be fitted by using the
same model used for the total spectrum and by accounting for the
variability with a variation of the normalization
of the transmitted powerlaw (i.e. the one dominating above 4 keV), i.e.
by ascribing the observed flux changes to intrinsic variability of the nuclear
source. The results of these two fits are summarized in Table 1 and in Fig. 3.

\begin{figure}[h]
\centerline{
\resizebox{\hsize}{!}{\includegraphics{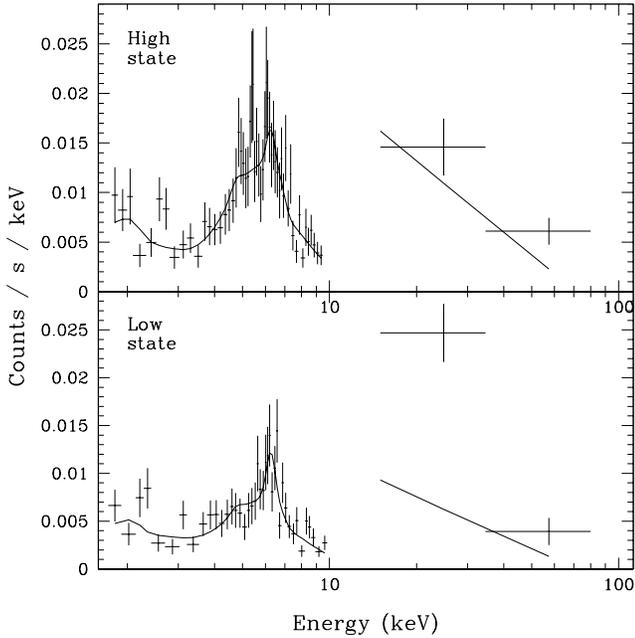}}}
\caption{{Spectra of NGC 1365 in its high
and low state (upper and lower panel respectively).
The models are obtained by a
fit to the 2-10 keV spectrum (excluding the PDS points).}}
\end{figure}

The light curve of the PDS behaves differently.
The variations in the 15--35 keV band are anticorrelated with
the variations in the 4--10 keV (MECS) spectral band (Fig. 3).
There are two possible explanations for this anti-correlation.
The Compton, cold reflection is most effective at $\sim 30$ keV, therefore
the anticorrelation could reflect a real delay between the innermost
primary source and the cold reprocessing material. However, this
scenario would require a 30 keV reflection efficiency of at least 60\% that is
very high, although not completely ruled out by models, depending on the
geometry of the reflector
(eg. Ghisellini et al. 1994). Alternatively, the observed
variability at 30 keV could be ascribed to the other Sy2 (NGC1386) in the
PDS beam.
These two interpretations are also supported by an analysis of Fig. 3, in which the
models are obtained by fitting the MECS data only. The extrapolation of these models
at higher energies fall short to account for the PDS data, in agreement with the
hypothesis of a contamination by an extra source or by a delayed, reflected
component.

The fact that the softer part of the spectrum is almost constant indicates that
the off-nuclear source does not contribute to the observed variability.
Indeed, the spectrum of the latter source is an unabsorbed power law
(Komossa \& Schulz 1998)
and, therefore, its
contribution to the variability should be significant in
the soft band (1.65-4 keV) too, in contrast to what observed.
The strength of the variability (the luminosity varied by 1.3$\times 10^{41}$
erg s$^{-1}$) also indicates that the
contribution of the off--nuclear source is marginal. The highest
known state of this source is that observed by ASCA in 1995, when the total
luminosity in the 2-10 keV band was $\sim 4\times 10^{40}$ erg s$^{-1}$,
that was already
an exceptional value for a non-nuclear galactic source.
Moreover, the measured flux of the soft component
is at the same level of that measured by ASCA.
Therefore, from both the spectral and time analysis we can reasonably 
assume that the emission of the off-nuclear source
during our observation was not significantly
 higher than during the ASCA observation.

\section{Discussion}
When comparing our BeppoSAX (1997) data  with the past ASCA (1994-95) data there are
differences that are not trivial to explain.

a) In 1994 the spectrum of NGC 1365 was dominated  by a (cold) reflected
component. The measured 2-10 keV flux was 1.1$\times10^{-12}$ erg cm$^{-2}$
s$^{-1}$. In 1997 we find a variable, direct component, absorbed by a
column density N$_H = 4.0\times 10^{23}$ cm$^{-2}$, whose 2-10 keV flux
is 6.6$\times 10^{-12}$ erg cm$^{-2}$ s$^{-1}$, i.e. 6 times higher than in
1994.

b) In the ASCA spectrum two iron emission lines are present, one at
E=6.4 keV (cold) and one at E=6.7 keV (warm).
In the SAX spectrum we also find evidence for a cold and a warm (E=7
keV) component of the
iron line, but the flux of the cold component is three
times higher than in 1994, while the warm component remains constant
within the statistical errors.

c) The energy of the cold iron line in the SAX spectrum (6.29 keV rest frame)
is lower than the expected value (6.4 keV).
As a consequence, the energy of the cold line in the SAX spectrum
is also lower than the cold (fainter) line observed in 1994, that is
consistent with 6.4 keV.\\

The model we propose to explain these data is based on a
multi-component absorber/reflector, composed by a warm, ionized component
in the sub-parsec scale, a cold molecular torus and 
another warm diffuse component outside the
torus. Fig. 4 schematically shows the various components of the model along
with their contribution to the observed spectrum.
In the following we discuss in detail each of these components and
spectral features.

\subsection{The cold absorber}
A cold absorbing medium is requested to explain the photoelectric
cutoff in the SAX spectrum, that is commonly observed in most obscured
Seyfert galaxies (Bassani et al. 1999). This medium is generally identified with
the obscuring molecular torus expected by the unified model
(Antonucci 1993).

\subsection{The cold iron line}
The cold iron line  at 6.4 keV
is thought to be emitted by the accretion disk and, in part,
by the circumnuclear torus predicted by the unified model (see eg.
Ghisellini et al. 1994 and Matt et al. 1991).
There are two possible explanations for the observed redshift of the iron
line to 6.29 keV in the BeppoSAX spectrum of NGC 1365:
a relativistic redshift (if the line is produced in the inner part
of an accretion disk) or a resonant absorption line at E$\simeq$6.6 keV
that shifts the center of the 6.4 keV line. We discuss in some detail each of these
two models in the following.

\begin{description}
\item[{\it Model 1.}]
We fitted the emission line with the standard DISKLINE model in the Xspec 10.0
code for spectral analysis.
In this model
the redshifted profile is due to the general relativistic
effects and the Doppler broadening.
All the parameters of the model (the inner
and outer radius of 
the disk and the inclination angle) were left free, except for the line energy,
which was frozen to E=6.365 keV (i.e. 6.4 keV rest frame). Details of the
line fit are given in Table 2a. The other parameters of the model are the
same as in Table 1, and their best fit values are equal, within the
errors, to those in Table 1.
The fit with the relativistic disk line is worse than the one
shown in Table 1, though it is still acceptable ($\Delta \chi^2=2$).

We note that the fit requires an angle between the disk axis and the line
of sight lower than 30 degrees (at the 90\% confidence level), i.e.
a disk oriented face-on. This geometry is not favored by the unified schemes,
since this object is characterized by an obscured nucleus (inferred both
from the optical spectrum and from the X-ray absorption) and, therefore,
the torus and the accretion disk are expected to be oriented edge-on.
However, a warped disk could solve this inconsistency. 

\item[{\it Model 2.}]
We now discuss the alternative model of the warm absorption Fe line.
A warm absorber in the central region of an AGN has been observed in
several Seyfert 1 galaxies
with a column density as high as several 10$^{23}$ cm$^{-2}$ 
(eg. Komossa \& Greiner 1999).
If this absorber is in an ionization state between Fe XXIV and Fe XIV
then the
resonant K$_\alpha$ transition can be both in emission and in
absorption at E$\approx$6.5--6.7 keV.
Matt (1994) predicts a Fe K$_\alpha$ resonant absorption line of EW of
20-30 eV for an ionized absorber with N$_H\simeq 10^{23}$
cm$^{-2}$, temperature T$\simeq 10^6 $ K, and a non-isotropic
spatial distribution around the central source.
The equivalent width can be larger if the temperature is higher
(T=10$^7$ K is an acceptable value for the region around the
accretion disk) and if the velocity dispersion of the warm absorber is high,
so that the broadening of the absorption line prevents its saturation.
For example, following Matt (1994),
if we assume T=10$^7$ K and a turbulence of $\sim$ 500
km s$^{-1}$ the EW of the absorption line can be $\sim$ 100 eV.
The combination of the cold emission line and the warm absorption line,
convolved at the spectral resolution of BeppoSAX, could result
in an emission line
whose center is apparently redshifted (Fig. 4).
A more quantitative description of this model  is given in the Appendix and
in Table 2b. 

\end{description}

\begin{table}
\centerline{\begin{tabular}{ccc}
\hline
Parameter&Best-fit value\\
\hline
&&\\
\multicolumn{2}{c}{\bf a) Relativistic line model} \\
Line energy & 6.365 (fixed)\\
Disk inner radius & R$_{IN}> 6 R_g$\\
Disk outer radius & unconstrained\\
Inclination angle & 18.4$^{+12}_{-18}$ deg\\
Line EW & 400$^{+300}_{-200}$ eV\\
$\chi^2$/d.o.f. &69/67\\
\hline
&&\\
\multicolumn{2}{c}{\bf b) Absorption line model} \\
Emission line Energy & 6.365 keV (fixed)&\\
Emission line EW & 400$^{+120}_{-200}$ eV\\
Absorption line Energy & 6.63$^{+0.25}_{-0.20}$ keV\\
Absorption line EW & 80 eV (fixed)\\
$\chi^2$/d.o.f. &66.3/68\\
\hline
\end{tabular}}
\caption{\footnotesize{Best fit parameters of the two models proposed to
explain the observed profile of the cold Fe emission lines, as discussed
in Sect. 3.2.}}
\end{table}

\subsection{The long term variability}

As outlined above, the X-ray emission and spectrum
of NGC 1365 is very different in
the two observations performed by ASCA and BeppoSAX. This behavior is
reminiscent of another well known case of similar long term variation,
i.e.
NGC 4051 (Guainazzi et al. 1998).

The differences between
the ASCA (1994) and SAX (1997) spectra, and in particular the
flux variation, can be explained in two scenarios: 1) a Compton thick cloud
(i.e. with N$_H > 10^{24}$ cm$^{-2}$) obscured
the nucleus in 1994 by passing through our line of sight, thus
making the 2--10 keV spectrum reflection dominated; alternatively 2)
the intrinsic emission of the active nucleus might have been
quiescent (or much reduced) in that period. The latter
case would be indistinguishable from the pure-reflection scenario,
because of the spectral similarity between a reflection spectrum and a
Compton-thin spectrum with N$_H\sim 4\times10^{23}$ cm$^{-2}$ when the
signal-to-noise is low (M98), as it is the case for the ASCA spectrum.
Moreover, in case 2) the observed emission
could be composed both by a direct and a reflected component. Also,
it is unlikely that the direct emission dominates, because a) in
this case some variability on short time scales would be expected, while
the ASCA light curve is constant within the errors (I97); and b) the
high equivalent width of the iron lines implies an highly efficient
reflection.

We cannot easily
distinguish between hypothesis 1) and 2), because both cases predict a
reflection dominated
spectrum, which depends only on the structure of the reflecting
medium. However, a very interesting result, regardless of which of the two
models applies, is that in both scenarios a high reflection efficiency is
required:
the ASCA 2-10 keV flux is 5.2\% of the SAX
N$_H$--corrected flux, that is near to the maximum possible reflection
efficiency,
according to theoretical models (Ghisellini
et al. 1994). According to these models the reflection efficiency is
strongly dependent on the column density of the reflecting material, and
is negligible for N$_H < 10^{24}$ cm$^{-2}$. We therefore conclude that
the reflection is not due to the same obscuring medium responsible for the
photoelectric cutoff observed in the SAX spectrum that,
according to our fits, has a column density (N$_H = 4\times 10^{23}$cm$^{-2}$)
much lower than what required to provide an efficient Compton reflection.
There are two simple models
 that could explain this discrepancy:
 \begin{itemize}
 \item the torus could be composed by a large number of thick clouds and
 diffuse gas with lower density and relatively low column density.
 Assuming this geometry the reflection efficiency could be high, and the
 SAX observation could have been performed when none of the thick clouds was
 intersecting our line of sight. However we note that in this scenario the
 covering factor of the clouds must be high, in order to make the
 reflection efficiency high enough. On the other hand,
 NGC 1365 was in a Compton thin state in two out of the three past
 observations (the Ginga and BeppoSAX ones), suggesting that the covering
factor of the thick clouds cannot be too high.

\item Alternatively, the obscuring torus might be characterized by a stratified
structure, with a column density in excess of $10^{24}$cm$^{-2}$ on the
equatorial plane and much lower on the edge, the latter being along our line of
sight.
This possibility is in agreement with some models that ascribe the
intermediate Seyfert classification to orientation
effects.
\end{itemize}

\begin{figure}
\centerline{
\resizebox{\hsize}{!}{\includegraphics{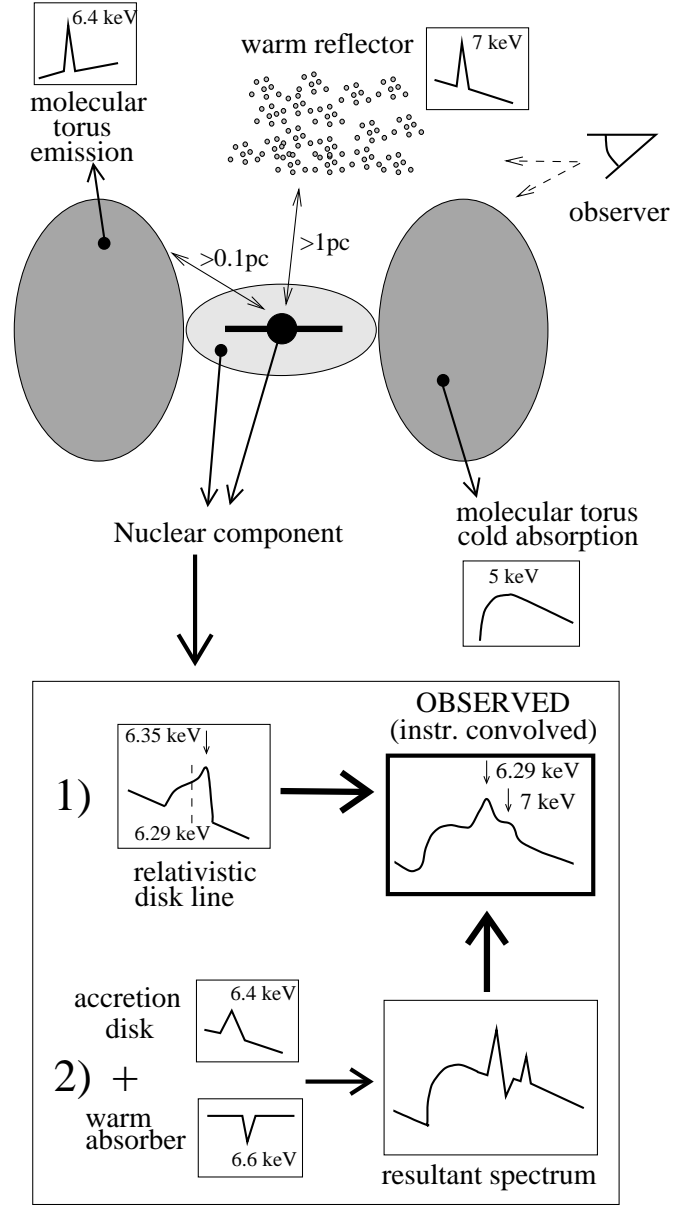}}}
\caption{{Sketch illustrating the geometry of the absorbing/reprocessing
material that we propose to explain the observed spectral components.
The observed profile of the cold iron line (6.29 keV rest
frame) can be reproduced by
both models 1) and 2).}}
\end{figure}

Finally, we note that the presence of a warm absorber in the central
region of the nucleus, speculated in Sect.3.2,
could also provide an explanation to the low--flux spectrum
measured by ASCA alternative to those discussed above:
a change in ionization state of the warm absorber could
introduce a much higher absorption that, henceforth, could be responsible
for the lower flux observed in the ASCA data.
Indeed, if the warm absorber is located close to the central source, as
required by our model, a decrease of the intrinsic luminosity could be
followed, with a short time delay, by a decrease of the ionization state
of the absorber and, as a consequence, by an increase of the absorption.
However, this effect is unlikely to provide all the additional column
density required from the ASCA data, basically for two reasons: 1) the
N$_H$ required ($\sim 10^{24}$ cm$^{-2}$) is much higher
than any previous measurement of warm absorbers
(as stated also in the Appendix, typical values of
N$_H$ for warm absorbers are several 10$^{22}$ cm$^{-2}$, with a few
cases of measured N$_H > 10^{23}$ cm$^{-2}$); 2)if a warm absorber with
N$_H\sim 10^{24}$ cm$^{-2}$ is present, we expect to detect a deep iron
absorption edge in the BeppoSAX spectrum, that is not observed. The
maximum {\it warm} N$_H$ for which the iron edge
is not detectable over the noise
(in excess to the Fe edge due to the cold absorber)
is a few 10$^{23}$ cm$^{-2}$.

\subsection{The cold mirrors}
 Whatever is the reason of the lower flux during the ASCA
observation, the comparison of the iron emission lines in the SAX and
ASCA spectra provides interesting constraints on the geometry and on the
efficiency of the reprocessing/ reflecting material.
The flux of the cold iron line in 1997 (SAX)
is three times higher than in 1994 (ASCA), 
confirming that the iron line is produced both by the obscuring torus and
by the reflection on
the accretion disk: when the nuclear source
is active (or visible)
 both components are detected, while when the nucleus is
inactive (or obscured) only the component reflected by the torus is detected.
If we assume that in 1994 the nucleus was in a low state (that is the 
most likely scenario, as discussed above) then we can constrain the size of
the torus that produces one-third of the cold iron line, by taking
advantage of reverberation limits. We do not have information about the
period when the nucleus first faded before the ASCA observation. However,
the ASCA observation was obtained in two parts
separated by 6 months and the two spectra are nearly identical, this
constrains the size of the cold reflecting torus to be larger than
6 light-months, i.e. $> 0.15$ pc, in agreement with other independent
estimates (Antonucci 1993, Gallimore et al. 1997, Greenhill \& Gwinn 1997).

\subsection{The warm mirror}
The warm iron line in AGNs is thought to be emitted by the
circumnuclear hot gas that is responsible for the reflection of the
(polarized) broad lines, i.e. the so-called {\it warm mirror} (Matt et
al. 1996).
To check if the discrepancy between the warm line energies in the
BeppoSAX and ASCA spectra is statistically significant, we retrieved and
analized the ASCA data from the ASCA public archive. We found that the
best fit energy of the warm line is E=6.85 keV (rest frame), and E=7 keV
is still acceptable ($\Delta \chi^2=1$). The line fluxes are also
compatible within the errors. This analysis suggests that 
the warm ionized reflector that produces this line
has not changed its state between tha ASCA (1994) and BeppoSAX (1997)
observations, and therefore should be
located at a distance larger than 3 light years from the nucleus.

Note that the warm reflector is  physically distinct from the
warm absorber responsible for the putative iron absorption system discussed
in Sect. 3.1, both because
of the different ionization properties and different size. As discussed in
Sect. 3.1, the warm absorber
must have a ionization state between FeXIV and FeXXIV,  while the warm mirror
emits a line at 7 keV that corresponds to FeXXV. Moreover, the ionized
absorber must be located in the vicinity of the black hole (around the
accretion disk?) so that turbulent velocity can broaden the absorption line
preventing its saturation, while the warm reflector must be located on scales
larger than 1 pc, as discussed above.

\subsection{The short term variability}
As discussed in Sect. 2.2,
the emission observed by BeppoSAX is strongly variable in the 4-10 keV
band. The light curve is very well fitted by a sinusoid with a period of
$\sim 45000 $ seconds. Even though our observation is too short to
justify any claim of periodicity, recent studies on the periodicity of
the Seyfert nucleus in IRAS 18325-5926 (Iwasawa et al.
1998) make this subject very interesting. Anyway, no conclusion can be
drawn without longer observations.

\section{Conclusions}
We presented new BeppoSAX data in the 0.1--100 keV range of the Seyfert 1.8
galaxy NGC1365.
The spectrum is characterized by a continuum absorbed by a cold gaseous
column density of $\rm N_H = 4\times 10^{23}$ cm$^{-2}$ and an iron K$\alpha$
emission complex that is well fitted by a cold component at 6.29 keV and
a warm component at 7 keV (rest frame). At energies below the absorption cutoff
(E $<$ 4 keV) a soft excess is present.

The cold absorption is probably due to the obscuring torus predicted by unified
model of AGNs. The continuum is strongly variable during the BeppoSAX
observation. The variability is mostly due to the hard component of the
spectrum above the photoelectric cutoff (4--10 keV), while the soft component
(1.65--4 keV) is essentially constant. The rapid variability very likely
reflects variations of the central engine. Instead, the soft excess is
probably due to an extended component, either associated to starburst activity
or to hot gas in the Narrow Line Region. 

The BeppoSAX spectrum is 6 times brighter than during two ASCA observations
of NGC~1365 taken about 3 years earlier. The latter spectra were characterized
by a flat continuum, indicative of cold Compton reflection, very likely from
the circumnuclear torus.

The high reflection efficiency, deduced from the comparison of the ASCA
and BeppoSAX spectra, requires a column density of the reflector much
higher than that measured in absorption. We conclude that the
circumnuclear medium is strongly inhomogeneous: the torus could contain
Compton thick clouds or, alternatively, has a steep density gradient
from the edge to the equatorial regions.

The fading of the direct emission during the ASCA
observations can be explained in two ways: the central
engine was hidden by a Compton thick cloud or, most probably, the nucleus
was in an intrinsically low state. In the latter scenario, the temporal
behavior of the cold and the warm iron lines indicate that the cold reflecting
torus must be located at a distance larger than 0.15 pc, while the warm
mirror must be located at a distance larger than 1 pc. Both the circumnuclear
torus and the accretion disk contribute to the emission of the cold Fe line,
in a proportion of about 1:2 respectively.

The cold iron line is significantly redshifted with respect to its nominal
value. More specifically we measure a line peak (rest frame) of
6.29 keV, that is inconsistent with the nominal value of 6.4 keV
at a significance level higher than 99\%. A disk relativistic line can fit
the observed profile, though the fit is worse than the analytical fit. Also,
according to this fit the accretion disk must be oriented face on, that is
an improbable geometry for an absorbed AGN like NGC 1365. Alternatively,
we propose that the shift of the cold iron line is caused by a warm absorber,
along the line of sight (with $\rm N_{warm}\approx 10^{23}$ cm$^{-2}$),
that introduces an absorption Fe line at 6.5--6.7 keV:
the combination of the cold emission line and the warm absorption line,
convolved with the spectral resolution of BeppoSAX, results in an emission line
whose center is apparently shifted at 6.29 keV. The spectral fit of the data
with this second model is significantly better with respect to the
relativistic disk line.

\begin{acknowledgements}
We thank the anonymous referee for useful comments.
G.R. and R.M. acknowledge the partial financial support
from the Italian Space Agency (ASI)
through the grant ARS--99--15 and from the
Italian Ministry for University
and Research (MURST) through the grant Cofin98-02-32.
\end{acknowledgements}

\appendix
\section{Details on the warm absorber model for the iron line
redshift}

In this Appendix we discuss more in detail the spectral fit and the
implications of the model of the warm iron absorption to explain the
redshift of the cold iron line described in Sect. 3.3 (model 2).

We fit our BeppoSAX data with a model whose components are the
same as in Table 1, except for the ``cold'' iron line at E=6.257 keV that
was replaced with a narrow line with energy frozen at 6.365 keV (6.4 keV rest
frame) and a
narrow absorption line with EW=80 eV and E$\sim$6.6 keV. Details of the
fit are given in Table 2b.
The best fit with this model
is better than in the case of the relativistic line model at high
statistical confidence ($\Delta \chi^2$=2.7 with one {\it additional}
degree of freedom).
Unfortunately, the statistics is not high enough to study the low and
the high state separately in the framework of this model and, in particular,
variations of the absorption line between the low--state and the
high--state: in both cases the best fit
value for the iron emission line energy
is lower than the canonical one, but the shift is significant only
at a $\Delta \chi^2 \sim 2$ level and therefore the absorption line
cannot be well constrained.

The redshift of the
Fe line could be simply due to random
fluctuations (the signal-to-noise in our spectrum
 is not very high). To check this possibility
we performed a simulation by means of the XSPEC 10 code
by using a very high integration time and with the same parameters as above
(without absorption).
After convolving with the response
matrix of the MECS instrument, we fitted the simulated spectrum
with a
single gaussian (in emission).
The best fit of the resulting spectrum is a gaussian at E=6.3 keV and EW=190
keV, in agreement with what
observed in
NGC 1365 (after correction for the continuum cold absorption), thus
confirming that the combination of the emission and the absorption lines
results in a redshifted observed lines and that the effect is not due
to the limited signal-to-noise.

Summarizing, a possible scenario to explain the Fe line profile in the BeppoSAX
spectrum is that a broad iron emission line is formed at
the surface of the central accretion disk (similarly to what is observed in
several Sy1s) and then it is partially absorbed by a warm circumnuclear
gas that causes an apparent redshift of the cold line centroid.

As discussed above, for the
absorption system to be effective in redshifting the centroid of the
cold
emission line the column density of the warm absorber must be
$N_{\rm warm} \simeq 10^{23}$cm$^{-2}$ or higher. 
Although observed in some Sy1s,
typically warm absorbers have column densities significantly lower
(Reynolds 1997).
Possibly, as illustrated in Fig. 4,
the warm absorber is preferentially distributed in the equatorial plane
of
the torus/disk system and, as a consequence, the edge--on lines of sight
(as it is probably
the case for NGC1365) are characterized by higher column densities
of the warm gas.

The warm absorption model is favored both because it fits better
the observed data and because the relativistic line model requires a
geometry
that is improbable for this object. However, the relativistic line model
cannot be rejected.

\end{document}